\documentclass[
reprint,
amsmath,amssymb,
aps,
]{revtex4-2}

\usepackage{graphicx}
\usepackage{dcolumn}
\usepackage{bm}
\usepackage{lipsum}

\usepackage{amssymb}
\usepackage[toc]{glossaries}
\usepackage{dsfont}
\usepackage{siunitx} 

\newlength{\revtexFigWidth}
\newlength{\revtexFigWidthDouble}
\makeglossaries

\newacronym{rc}{RC}{reservoir computing}
\newacronym{ls}{LS}{local states}
\newacronym{gls}{GLS}{generalized local states}
\newacronym{ml}{ML}{machine learning}
\newacronym{cc}{CC}{cross-correlation}
\newacronym{mi}{MI}{mutual information}
\newacronym{sm}{SM}{similarity measure}
\newacronym{l96}{L96}{Lorenz-96}
\newacronym{ks}{KS}{Kuramoto-Sivashinsky}
\newacronym{mle}{MLE}{maximal Lyapunov exponent}
\newacronym{ann}{ANN}{artificial neural network}
\newacronym{rnn}{RNN}{recurrent neural network}
\newacronym{ffn}{FFN}{feed forward network}
\newacronym{pdf}{PDF}{probability distribution function}
\newacronym{nrmse}{NRMSE}{normalized root mean square error}
\newacronym{sn}{SN}{spatial neighborhood}
\newacronym{edtrk4}{ETDRK4}{exponential time-differencing fourth-order Runge-Kutta}
\newacronym{os}{OS}{orbit separation}

\begin{document}

\title{Predicting high-dimensional heterogeneous time series employing generalized local states}

\author{Sebastian Baur$^{1,2}$}
\author{Christoph R\"ath$^{1}$}
\affiliation{
$^{1}$Institut f\"ur Materialphysik im Weltraum, Deutsches Zentrum für Luft-und Raumfahrt, M\"unchner Stra{\ss}e 20, 82234 We{\ss}ling, Germany and\\
$^{2}$Fakult\"at f\"ur Physik, Ludwig-Maximilians-Universit\"at M\"unchen, Schellingstra{\ss}e 4, 80779 Munich, Germany
}

\date{April 11, 2021}

\begin{abstract}

We generalize the concept of \gls{ls} for the prediction of high-dimensional, potentially mixed chaotic systems. The construction of \gls{gls} relies on defining distances between time 
series on the basis of their (non-)linear correlations. 
We demonstrate the prediction capabilities of our approach based on the
\gls{rc} paradigm using the \gls{ks}, the \gls{l96} and a combination of both systems. 
In the mixed system a separation of the time series belonging to the two different systems is made possible with \gls{gls}. More importantly, prediction remains possible with \gls{gls}, where the \gls{ls} approach must naturally fail.  
Applications for the prediction of very heterogeneous time series with \gls{gls}s are briefly outlined.  

\end{abstract}

\maketitle

\section{Introduction} 
Tremendous advances in predicting the short and long term behavior of complex systems have been made in recent years by applying 
machine learning
~\cite{Lu2018, Vlachas2018, Pathak2018, Chattopadhyay2020, Mabrouk2020a, Haluszczynski2020a, Carroll2020}.
reservoir computing
turned out to be a very promising 
machine learning 
approach
as it combines outstanding performance with conceptual advantages like very fast and comparably transparent learning and possibly appealing hardware realizations of 
reservoir computing
~\cite{Bompas2020, Marcucci2020}.

For high-dimensional systems 
reservoir computing
suffers like other 
machine learning
methods from the "curse of dimensionality" meaning that the number of nodes of the network representing the reservoir has to be considerably larger than the dimensionality of the input data rendering the training unfeasible with a naive 
reservoir computing
approach. With a parallel prediction scheme based on 
local states
~\cite{Parlitz2000}, however, the forecasting of high-dimensional chaotic spatiotemporal systems of arbitrarily large extent becomes possible~\cite{Pathak2018a, Zimmermann2018}.

The definition of 
local states
relies on defining spatial local neighborhoods for each time series to be predicted. Thus, the knowledge of the position of the time series in space is a necessary prerequisite for defining 
local states.

The similarity of time series can be defined in a much more general way by deducing a distance measure and thus a local neighborhood from the correlations among the time series~\cite{Mantegna1999}.
Those generalized similarities led -- among others -- to a reasonable, fully data-driven taxonomy of the stock market~\cite{Onnela2003} while crucial differences between the linear and nonlinear correlation structure of the stock market especially during crises were reported later~\cite{Haluszczynski2017}.

In this Article, we employ this approach to define 
generalized local states
for the prediction of high dimensional systems with which some of the shortcomings of the 
local states
approach can be overcome.

\section{Systems and Simulation Details}
For modeling high dimensional, spatiotemporal, chaotic systems, the \gls{l96} and \gls{ks} systems have become widely used in the \gls{rc} community~\cite{Pathak2018, Lu2017, Vlachas2020, Huang2020}.
The \gls{l96}~\cite{Lorenz1996,Wilks2005} system is defined as
\begin{equation}
\frac{d x_{j}}{d t}=\left(x_{j+1}-x_{j-2}\right) x_{j-1}-x_{j}+F,
\end{equation}
where $\boldsymbol{x}(t)$ is the $D$-dimensional state vector of the system and $F$ the forcing parameter. 
In this study, we set $D=40$ and $F = 5$ resulting in a chaotic system for which we calculate a \gls{mle} of $\Lambda_{max} = 0.45$ using Sprott's method of \gls{os}~\cite{sprott2003chaos}. 
We use the fourth-order Runge–Kutta method~\cite{press1992numerical} for the simulation, with a time step of $\Delta t = 0.05$.

The second model, widely used to model a variety of weakly turbulent fluid systems~\cite{kuramoto1976persistent, pomeau1985kuramoto}, is the \gls{ks} system~\cite{Kuramoto1976}. 
Its PDE reads
\begin{equation}
\partial_{t} u+\partial_{x}^{4} u+\partial_{x}^{2} u+u \partial_{x} u=0\,,
\end{equation}
where the field $u(x, t)$ is defined on some domain size $L$.
In this study, we use a domain size of $L = 22$ with periodic boundary conditions
$u(x+L, t)=u(x, t) \quad \text{ for all } 0 \leq x \leq L$.
For the numerical treatment, the equations are discretized on a grid of $D = 40$ points, the same size as the \gls{l96} system, and numerically integrated, with a time step of $\Delta t = 0.5$, using the fourth order time-stepping method ETDRK4~\cite{Kassam2005}. Using \gls{os} we find a \gls{mle} of $\Lambda_{max} = 0.049$.

\section{Reservoir Computing}
We use a standard setup for reservoir computing.
At the center of \gls{rc} is a network, in the following called the reservoir $\mathbf{A}$, created as a sparse random $D_{r} \times D_{r}$ network with average node degree $\kappa$.
After network generation, all its random, uniformly distributed connection strengths are scaled to have a predetermined fixed spectral radius $\rho$.
The $D_{\text{in}}$ dimensional input $\boldsymbol{x}(t)$ interacts with the reservoir state $\boldsymbol{r}(t)$ through an input coupler which in our case is a sparse $D_{r} \times D_{\text{in}}$ matrix $\mathbf{W}_{\text{in}}$.  Following~\cite{Lu2018}, $\mathbf{W}_{\text{in}}$ is created such that one element in each row is chosen uniformly between $[-\omega, \omega]$ where $\omega$ is the input coupler scaling parameter. All other elements of $\mathbf{W}_{\text{in}}$ are zero. The input data then connects with the reservoir state $\boldsymbol{r}(t)$ of the previous time step via activation function $\tanh(\cdot)$ to advance the reservoir state by one step in time
\begin{equation}
\boldsymbol{r}(t+\Delta t)=\tanh \left(\mathbf{A} \boldsymbol{r}(t)+\mathbf{W}_{\text {in }} \boldsymbol{x}(t)\right).
\end{equation}
 
We choose a simple matrix as output coupler $\mathbf{W}_{\text{out}}$, whose elements are  determined in the training phase via ridge regression 
\begin{equation}
\mathbf{W}_{\text{out}}=\min_{\mathbf{W}_{\text{out}}}\left(\left\|\mathbf{W}_{\text{out}} \boldsymbol{\tilde{r}}(t)- \boldsymbol{y}_{\text{T}}(t)\right\|+\beta\left\|\mathbf{W}_{\text{out}}\right\|\right),
\end{equation}
where $\boldsymbol{y}_{\text{T}}(t)$ is the $D_{\text{out}}$ dimensional target output. $\boldsymbol{\tilde{r}}$ is a nonlinear transformation of the reservoir state $\boldsymbol{r}$ here chosen to be $\boldsymbol{\tilde{r}} = [\boldsymbol{r}, \boldsymbol{r}^2]^T = [r_1, r_2, ..., r_{D_{r}},  r_1^2, r_2^2, ...  r_{D_{r}}^2]^T$. This blow-up of the reservoir states first introduced by Lu et al.~\cite{Lu2017} actually serves to break the symmetries in the reservoir equations~\cite{Herteux2020a}. 

To avoid that the arbitrary initial state of the reservoir influences the regression results, training only starts after a washout phase of 20000 time steps.

Once trained, the output $\boldsymbol{y}(t)$ can be calculated from the reservoir states $\boldsymbol{r}(t)$ as $\mathbf{y}(t)= \mathbf{W}_{\text {out}}\boldsymbol{\tilde{r}}(t)$.
When using \gls{rc} for prediction, it can then be run autonomously by using the prediction of the previous time step $\boldsymbol{y}(t)$ as the input $\boldsymbol{x}_{\text{pred}}(t)$ to calculate the next predicted time step $\boldsymbol{y}(t+\Delta t)$. In this case one finds 
\begin{equation}
\boldsymbol{r}(t+\Delta t) =\tanh \left(\mathbf{A} \boldsymbol{r}(t)+\mathbf{W}_{\text {in }} \boldsymbol{x}_{\text{pred}}(t)\right).
\end{equation}

\section{Generalized Local States}
\subsection{Generalizing Local States}
\gls{gls} is based on the \gls{ls} approach proposed by Parlitz et al.~\cite{Parlitz2000} and used for RC prediction by Pathak et al.~\cite{Pathak2018}. While non-local implementations of \gls{rc} algorithms use just one network to process all input data, \gls{ls} and \gls{gls} partitions the input data into multiple subsets of smaller dimension each with their own reservoir. The reservoirs are then trained on and predict these subsets only. This provides an effective workaround for the curse of dimensionality by essentially parallelizing the prediction of one high-dimensional dataset using many lower-dimensional subsets.
These subsets are in the following called \textit{neighborhoods}. 
Each neighborhood itself consists of a number of \textit{core} variables and \textit{neighbor} variables and is assigned its own reservoir. Each reservoir in question then uses only the variables of the input making up its neighborhood to predict its core variables as accurately as possible. 

\begin{figure}[htp]
\centering
\includegraphics[width=\revtexFigWidth]{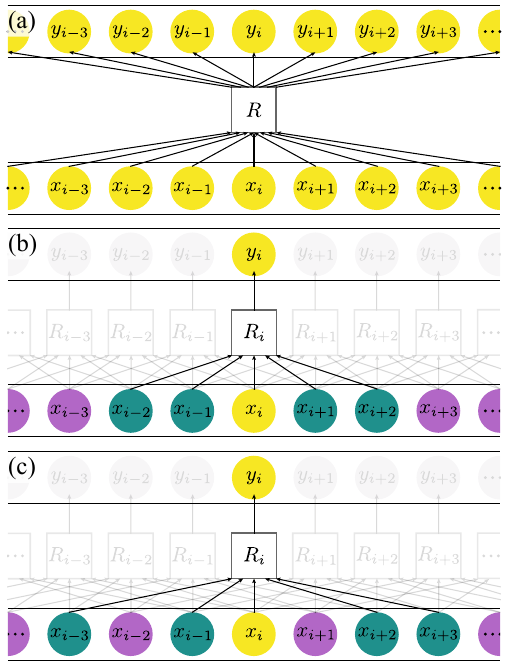}
\caption{\label{fig:fig_gen_loc_state_schematic} Schematic depicting different neighborhoods and associated data flow. Yellow circles mark the core variables, green the neighbors and purple all other variables of the full input. 
\textbf{(a)}
The neighborhood of simple, non-local RC. All dimensions of the input vector $\boldsymbol{x}(t)$ are core dimensions of the reservoir $R$. No neighbors or other reservoirs exist.
\textbf{(b)}
LS RC. The highlighted neighborhood of the reservoir has one core variable $x_i$ and four locally adjacent neighbors. These five variables are used as input for the reservoir $R_i$ to predict the core variables future state $y_i$. Many more reservoirs exist, each with their own neighborhood.
\textbf{(c)}
GLS RC. As in \textbf{(b)}, the neighborhood has one core variable and four neighbors, with the crucial difference being that the neighbors do not have to be locally adjacent to the core.}
\end{figure}

As such, the input time series for the reservoir assigned to the $i$-th neighborhood is not the full $D_{\text{in}}$ dimensional input time series $\boldsymbol{x}$ anymore, but instead a slice unique to this neighborhood $\boldsymbol{x}^{i}$ of dimension $D^i_{\text{in}} \leq D_{\text{in}}$.
Similarly, the trained output of the $i$-th neighborhood $\boldsymbol{y}^{i}$ is given by just its core variables and hence is an even smaller subset of the neighborhood of dimension $D^i_{\text{out}} \leq D^i_{\text{in}}$. These different neighborhoods are depicted in Figure~\ref{fig:fig_gen_loc_state_schematic}.

Between each prediction step the neighborhoods need a new input which, as typically $D^i_{\text{out}} < D^i_{\text{in}}$, can only come from the other neighborhoods' prediction. Hence between each prediction step, a new input vector $\boldsymbol{x}_{\text{pred}}$ is formed by the combined core variable prediction of all neighborhoods. As a consequence, each variable of the original input must be in one and only one neighborhood as a core variable. 

Pathak et al. originally introduced their \gls{ls} approach in the context of the \gls{ks} system, a purely locally interacting system. As such, their neighborhoods were chosen to have only spatially adjacent cores surrounded by contiguous buffer regions of neighbors (see Figure~\ref{fig:fig_gen_loc_state_schematic}b).

For systems where such a local interaction is not present, this scheme cannot be used. Nonetheless, the idea of locality in the sense of importance to a prediction is generalizable to something which we will call \textit{similarity}. As the choice of neighborhoods is essentially arbitrary, this allows the creation of neighborhoods not by which variables are locally closest to the core variables, but by instead including the variables most \textit{similar} to the cores as neighbors. Such a non-local \gls{gls} neighborhood is shown in Figure~\ref{fig:fig_gen_loc_state_schematic}c).

The effectiveness of this procedure of course depends on an appropriate choice of neighborhoods, such that the information necessary to predict their core variables is present in their (non-local) neighbors.

\subsection{Similarity Measures}
Our first measure to estimate the similarity of the time series is the (linear) \gls{cc} coefficient 
\begin{equation}
C_{X_{i}, X_{j}}=\frac{\sum_{t=1}^{n}\left(x_{i, t}-\bar{x}_{i}\right)\left(x_{j, t}-\bar{x}_{j}\right)}{\sqrt{\sum_{t=1}^{n}\left(x_{i, t}-\bar{x}_{i}\right)^{2}} \sqrt{\sum_{t=1}^{n}\left(x_{j, t}-\bar{x}_{j}\right)^{2}}}\,,
\end{equation}
$C_{x_{i}, x_{j}}$, where $x_{i}$ is the $i$-th variable of the time series $\boldsymbol{x}$.
To transform the \gls{cc} into a \gls{sm} we take its absolute value 
\begin{equation}
\text{SM}_{\text{CC}}(X_{i}, X_{j}) = \left| C_{X_{i}, X_{j}} \right| \in [0, 1],
\end{equation}
and associate a larger value as more similar. As such, both a high correlation as well as a high anti-correlation correspond to a high similarity.

Considering that chaotic time series are by their very nature nonlinear, we use one more measure that captures these nonlinear relationships, the \gls{mi}.
As we are working with long ($10^5$ time steps), well behaved time series, we will implement the widely used binning method~\cite{Shannon1948, Kraskov2004, Haluszczynski2017} as estimator for the \gls{mi}.
Akin to~\cite{Haluszczynski2017} we find empirically that for our time series of length $T=10^5$ a choice of $\left \lfloor{\sqrt{T/4}}\right \rfloor = 158$ bins of equal size works well.
We normalize the \gls{mi} as described by Strehl et al.~\cite{Strehl2003} leading to our \gls{mi} \gls{sm} 
\begin{equation}
\text{SM}_{\text{MI}}(X_{i}, X_{j})=\frac{I\left(X_{i}, X_{j}\right)}{\sqrt{H\left(X_{i}\right) H\left(X_{j}\right)}} \in [0, 1],
\end{equation}
where $I\left(X_{i}, X_{j}\right)$ is the \gls{mi} while $H\left(X_{i}\right)$ and $H\left(X_{j}\right)$ are the Shannon entropies of $X_{i}$ and $X_{j}$ respectively.

\subsection{Neighborhood Definitions}
Once a \gls{sm} has been chosen and calculated for all variables of the full input data, one needs to use it to create the neighborhoods. As the prediction output of each neighborhood is only given by its core variables, predicting these core variables as accurately as possible is most important. While a variety of choices are possible, we restrict ourselves to associate each neighborhood with exactly one core variable.
In the LS case, the neighbors of the core are simply the variables spatially closest to that core. We will call this a \gls{sn}.
In the GLS case, the exact analog is possible, where the neighbors of the core are the variables most similar to it as defined be the \gls{sm}.
We call the corresponding neighborhoods \gls{cc} or \gls{mi} neighborhoods, depending on the similarity measure used.

Lastly, it should be emphasized that the \gls{gls} method is in principle independent of not only the specific \gls{sm} used, but also of the chosen prediction method. Many other choices for the \gls{sm}, e.g. the transfer entropy, or the network, e.g. LSTMs, are conceivable.

\section{Experiments}
\subsection{Methods and Parameters}
To enable a fair comparison of the different neighborhood generation methods, the \gls{rc} hyperparameters are optimized for the \gls{ls} neighborhoods 
and then copied for the \gls{gls} neighborhoods without further adjustments.
Furthermore, we added noise to all training data following Vlachas et al.~\cite{Vlachas2018}. Without this noise we found the short term prediction accuracy to often be higher, but at the cost of an increased rate of failed realizations, and lower quality long term predictions in general. 
The added noise proved decisive in minimizing variance between network realizations and reducing the number of failed realizations, especially for the \gls{l96} system.
Heuristically, we found normally distributed noise with standard deviation $\sigma_{\text{noise}} =  1\%\,\sigma_{\text{data}}$, where $\sigma_{\text{data}}$ is the standard deviation of the training data, to be a sweet spot.
The hyperparameters used for all \gls{rc} are given in table~\ref{tab:rc_hyperparameters}.
\begin{table}[h]
\begin{center}
\begin{tabular}{ |c c c|  } 
\hline
 reservoir dimension & $D_{r}$ &  $5000$ \\ 
 average node degree & $\kappa$ & $3$\\
 spectral radius & $\rho$ & $0.5$\\
 input coupler scaling & $\omega$ & $0.5$\\
 ridge regression parameter & $\beta$ & $10^{-6}$\\
 noise level & $\alpha$ & $1\%$\\
 \hline
\end{tabular}
\caption{\gls{rc} hyperparameters used throughout this article.}
\label{tab:rc_hyperparameters}
\end{center}
\end{table} 

Transient effects of the simulated data were discarded before any synchronization, training or prediction took place. 
Similarly, each reservoir training and prediction is preceded by $2000$ synchronization steps.
All reservoirs were trained for a total of $10^5$ time steps using the noisy training data.

To calculate the \glspl{sm} we use the same noisy data used to train the reservoirs. 
As a result all neighborhoods shown here are representative examples, but not uniform for all realizations.
The neighborhoods are calculated for a single core and 18 neighbors, in the case of \gls{sn} and \gls{mi} neighborhoods, and 28 neighbors, in the case of a \gls{cc} neighborhood. 

These fixed neighborhood sizes for \gls{sn} and \gls{mi} were chosen to be similar to Pathak’s original paper~\cite{Pathak2018} which we found to be a good compromise between computation speed (small neighborhoods) and prediction accuracy (large neighborhoods).
While for them this results in good predictions, for the \gls{cc} \gls{sm} this leads to essentially all predictions diverging. 
This is likely the result of the linear \gls{cc} \gls{sm} not recognizing the importance of the core's nearest neighbors in these systems. As such, the \gls{cc} neighborhood size was increased to a total size of 29, the minimum where no predictions diverged.
\gls{cc} neighborhoods of different sizes for the \gls{l96} system are discussed in appendix~\ref{cc_neighborhood_size}.

\subsection{Predicting the L96 System}
\begin{figure}[tph]
\includegraphics[width=\revtexFigWidth]{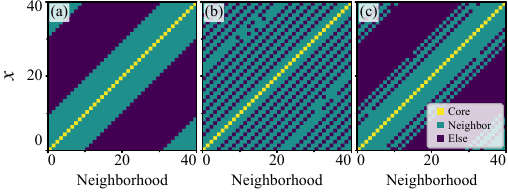}
\caption{\label{fig:L96_neighboorhoods} Example Neighborhoods for the \gls{l96} system. Each row depicts one variable of the time series, while each column represents a neighborhood. Each neighborhood is defined by the core variable (yellow) and its neighbors (green). 
\textbf{(a)}
\gls{sn} neighborhoods. The \gls{sn} neighborhood consists only of a core and its 18 nearest neighbors.
\textbf{(b)}
\gls{cc} neighborhoods. Each \gls{cc} neighborhood includes 28 neighbors chosen with the \gls{cc} \gls{sm}.
\textbf{(c)}
\gls{mi} neighborhoods with 18 neighbors.}
\end{figure}

Example neighborhoods for the  \gls{l96} systems are shown in Figure~\ref{fig:L96_neighboorhoods}.

While the \gls{sn} neighborhoods are simply defined as a single core and its nearest neighbors, the \gls{cc} and \gls{mi} neighborhoods warrant a closer look.
First and foremost, even though the \gls{mi} neighborhoods were calculated dynamically, without directly using the knowledge of \gls{l96} being a locally interacting system, the resulting \gls{mi} neighborhoods closely resemble the \gls{sn} neighborhoods. The \gls{cc} neighborhoods in contrast include many variables spatially far away from the core.

100 distinct random network realizations are generated for each system-\gls{sm} combination.
They are then trained 
and used to predict the same 300 sections of 10 Lyapunov times
length on the chaotic attractor of the \gls{l96} and \gls{ks} systems. 

\begin{figure}[tph]
\includegraphics[width=\revtexFigWidth]{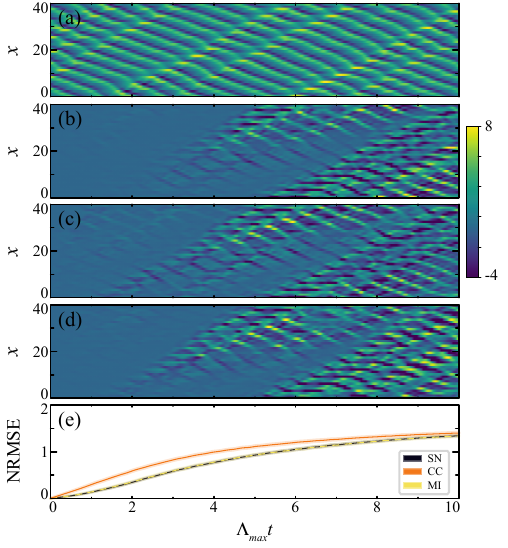}
\caption{\label{fig:L96_prediction_comparison}\gls{l96} system short term prediction comparisons.
\textbf{(a)}
Simulated data of the L96 system. 
\textbf{(b-d)}
Exemplary difference of the \gls{rc} prediction to the simulated data when using the \textbf{(b)} \gls{sn}, \textbf{(c)} \gls{cc} and \textbf{(d)} \gls{mi} neighborhoods.
\textbf{(e)}
\gls{nrmse} of \gls{sn}, \gls{cc}, and \gls{mi} prediction data averaged over first the 300 predicted sections and then the 100 network realizations. The error bands correspond to the $3\sigma$ standard deviation of the random network realizations.
We multiply $t$ by the \gls{mle} $\Lambda_{\text{max}}$ of the model, so that each unit on the horizontal axis represents one Lyapunov time.}
\end{figure}

To quantify the short term prediction accuracy, we define the \gls{nrmse} as
\begin{equation}
\operatorname{NRMSE}(\hat{\boldsymbol{y}})=\frac{\sqrt{(\hat{\boldsymbol{y}}-\boldsymbol{y})^{2}}}{y_{\text{max}} - y_{\text{min}}},
\end{equation}
where $\hat{\boldsymbol{y}} \in\mathbb{R}^D $ is the prediction at a single time-step, $\boldsymbol{y} \in\mathbb{R}^D $ the true signal and $y_{\text{max}}$ ($y_{\text{min}}$) the largest (smallest) value taken of any variable in the simulated data set.
Short term prediction results are shown in Figure~\ref{fig:L96_prediction_comparison}.

Looking at the short term predictions of the \gls{l96} system, it is very striking that the averaged \gls{nrmse} of the \gls{sn} and \gls{mi} neighborhoods coincide more or less exactly, while the \gls{cc} prediction is significantly worse. This performance drop is likely the result of the \gls{cc} neighborhood's inclusion of many variables which are far from the, ostensibly most important, close region around the core (see Figure~\ref{fig:L96_neighboorhoods}b).

It should be noted that the statistical stability that has been assessed here for the first time is remarkable. 
Often \gls{rc} algorithms exhibit a much larger spread of prediction qualities between different random networks~\cite{Haluszczynski2019}. This stability is partly attributable to finding the correct hyperparameters for the systems at hand, as suboptimally chosen hyperparameters often leave the best predictions intact while increasing the variance towards completely failed predictions drastically~\cite{Haluszczynski2019, Vlachas2020}.
However, empirically we found that the role of the noise added to the training data in achieving this stability, especially for the \gls{l96} system, cannot be understated either.

Using the \gls{os} method we can calculate the \glspl{mle} of the predictions as another characteristic to quantify long term prediction accuracy. For this purpose we let each of our reservoir realizations predict three distinct sections of the system attractor each 1000 Lyapunov times in length which are then used to calculate the \glspl{mle} shown in table~\ref{tab:40_nbhds_lyapunov_exponents_just_l96}.
\begin{table}[tph]
\begin{center}
\begin{tabular}{ |c|c|c|c|c| } 
\hline
System & SIM & \gls{sn} & \gls{cc} & \gls{mi} \\ 
\hline
\gls{l96} & $0.45$ & $0.43 \pm 0.01$  & $0.47 \pm 0.02$ & $0.44 \pm 0.02$ \\ 
\hline
\end{tabular}
\caption{\glspl{mle} calculated via the \gls{os} method for the simulated (SIM) and predicted trajectories using the \gls{sn}, \gls{cc} and \gls{mi} neighborhoods for the \gls{ks} and \gls{l96} data. The errors represent the $1\sigma$ standard deviation between network realizations.}\label{tab:40_nbhds_lyapunov_exponents_just_l96}
\end{center}
\end{table}

The \glspl{mle} of all three neighborhoods agree well with the \glspl{mle} calculated directly from the simulations. 
The details of the \gls{os} procedure used are described in appendix~\ref{orbit_separation}.
 
\subsection{Predicting the KS System}
\label{ks_prediction_analysis}
In preparation for the non-local system discussed in the following section, we predict and analyze the \gls{ks} system as we have just done for the \gls{l96} system.

\begin{figure}[!h]
\includegraphics[width=\revtexFigWidth]{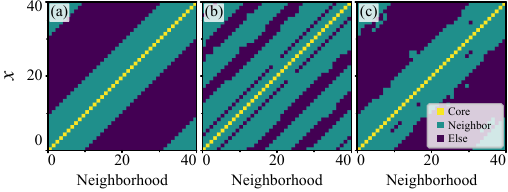}
\caption{\label{fig:KS_neighboorhoods} Example neighborhoods for the \gls{ks} system. Each row depicts one variable of the time series, while each column represents a neighborhood. Each neighborhood is defined by the core variable (yellow) and its neighbors (green). 
\textbf{(a)}
SN,
\textbf{(b)}
\gls{cc},
\textbf{(c)}
\gls{mi} neighborhoods.
}\end{figure}

\begin{figure}[!ht]
\includegraphics[width=\revtexFigWidth]{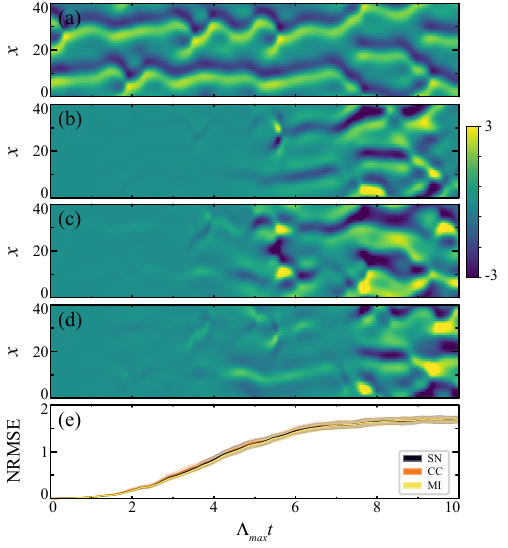}
\caption{\label{fig:KS_prediction_comparison}\gls{ks} system short term prediction comparisons.
\textbf{(a)}
Simulated \gls{ks} data. 
\textbf{(b-d)}
Exemplary error in the \gls{rc} prediction when using the \textbf{(b)} \gls{sn}, \textbf{(c)} \gls{cc} and \textbf{(d)} \gls{mi} neighborhoods.
\textbf{(e)}
\gls{nrmse} of \gls{sn}, \gls{cc}, and \gls{mi} prediction data averaged over first the 300 predicted sections and then the 100 network realizations. The error bands correspond to the $3\sigma$ standard deviation of the random network realizations.}
\end{figure}

Example neighborhoods and short term prediction results for the \gls{ks} system are shown in Figure~\ref{fig:KS_neighboorhoods} and Figure~\ref{fig:KS_prediction_comparison} respectively.

Looking at the short term predictions of the \gls{ks} system shown in Figure~\ref{fig:KS_prediction_comparison}, it is striking that the averaged \gls{nrmse} of all three neighborhoods coincide more or less exactly. This is remarkable especially when compared to the \gls{l96} system in which the \gls{cc} neighborhoods performed significantly worse.

Using \gls{os} to calculate the \gls{mle} we find them again to agree excellently with the simulated \gls{mle} as depicted in table~\ref{tab:40_nbhds_lyapunov_exponents_just_ks}.
\begin{table}[h]
\begin{center}
\begin{tabular}{ |c|c|c|c|c| } 
\hline
System & SIM & \gls{sn} & \gls{cc} & \gls{mi} \\ 
\hline
\gls{ks} & $0.049$ & $0.046 \pm 0.004$ & $0.048 \pm 0.001$ &  $0.048 \pm 0.001$  \\ 
\hline
\end{tabular}
\caption{
Same as table~\ref{tab:40_nbhds_lyapunov_exponents_just_l96} but for the \gls{ks} instead of the \gls{l96} system.
}
\label{tab:40_nbhds_lyapunov_exponents_just_ks}
\end{center}
\end{table}

\subsection{Predicting a non-local System}
\begin{figure}[tph]
\includegraphics[width=\revtexFigWidth]{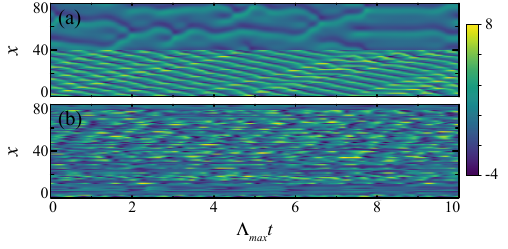}
\caption{\label{fig:L96_KS_before_and_after_shuffle_v1_min} Concatenated System
\textbf{(a)}
Simulated data combined from the \gls{l96} (variables 1-40) and \gls{ks} (variables 41-80) systems.
\textbf{(b)}
Simulated data as in \textbf{(a)} with the variables shuffled.
The time axis is scaled by the \gls{mle} $\Lambda_{\text{max}}$ of the \gls{ks} model.
}\end{figure}

To test the usefulness of \gls{gls} for non-locally interacting systems, we use the \gls{ks} and \gls{l96} systems to artificially create such a non-locally interacting test system. As depicted in Figure~\ref{fig:L96_KS_before_and_after_shuffle_v1_min} we do this by concatenating both systems and then randomly shuffling the 80 variables of the combined system.

As this combined system now is a composite of two systems with different time steps, it does not have a well defined Lyapunov exponent any more. For the sake of consistency we nonetheless continue the time axis rescaling in terms of Lyapunov exponents. To do so we use the larger of the two system's time steps per Lyapunov time as calculated in table~\ref{tab:system_lyap_properties}, hence at worst slightly underestimating our short term prediction results.
 
\begin{figure}[tph]
\includegraphics[width=\revtexFigWidth]{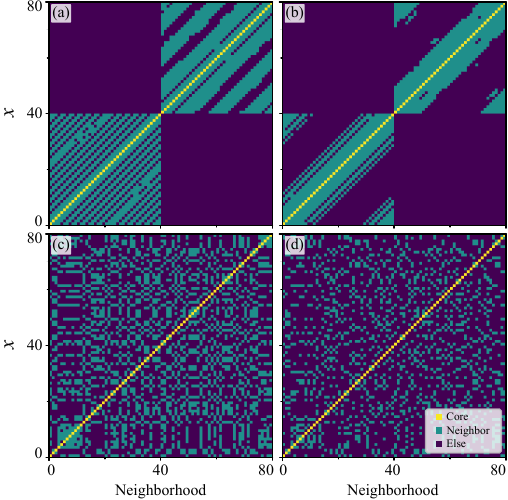}
\caption{\label{fig:L96_KS_neighboorhoods}Neighborhoods of the concatenated \gls{l96} and \gls{ks} systems. 
\textbf{(a)} is the \gls{cc} and \textbf{(b)} the \gls{mi} neighborhoods for the concatenated, un-shuffled system. Variable 1-40 of the data used to compute these neighborhoods come from the \gls{l96} system with variables 41-80 originating from the \gls{ks} simulation
In \textbf{(c)} the \gls{cc} and \textbf{(d)} the \gls{mi} neighborhoods for the shuffled system are shown.}
\end{figure}

As before, we can also calculate the \gls{sn}, \gls{cc} and \gls{mi} neighborhoods for this new concatenated system. The \gls{cc} and \gls{mi} neighborhoods are shown in Figure~\ref{fig:L96_KS_neighboorhoods}.

The \gls{sn} neighborhoods have been omitted from this Figure as they are, by definition, always the same. Fascinatingly, both the \gls{cc} and \gls{mi} neighborhoods in the combined but not shuffled systems look like they are composed of the individual system's neighborhoods. In fact, exactly this is the case as both the \gls{cc} and the \gls{mi} \glspl{sm} are able to completely separate the \gls{ks} and \gls{l96} systems. In the case of experimental data, such an analysis of the similarity structure can be highly informative as it exposes the underlying relationships between dimensions. 

\begin{figure}[tph]
\includegraphics[width=\revtexFigWidth]{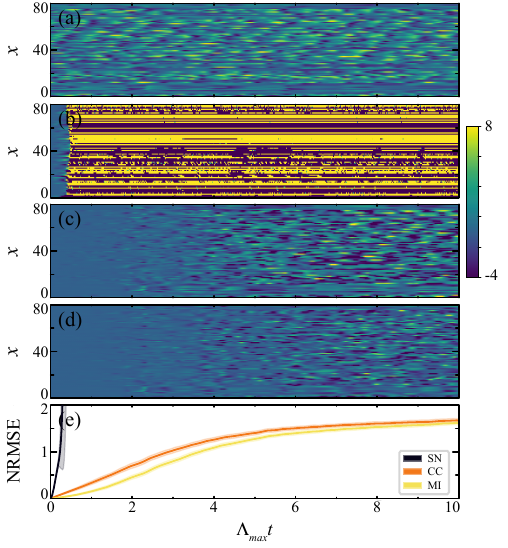}
\caption{\label{fig:L96_KS_shuffle_prediction_comparison} Shuffled system short term prediction comparison.
\textbf{(a)}
Simulated data of the combined shuffled KS-L96 system. 
\textbf{(b-d)}
Exemplary difference of the \gls{rc} prediction to the simulated data when using the \textbf{(b)} \gls{sn}, \textbf{(c)} \gls{cc} and \textbf{(d)} \gls{mi} neighborhoods. The color scale of the diverging prediction was cut to the ensure legibility of the other plots
\textbf{(e)}
\gls{nrmse} of \gls{sn}, \gls{cc}, and \gls{mi} prediction data averaged over first the 300 predicted sections and then the 100 network realizations. The error bands correspond to the $3\sigma$ standard deviation of the random network realizations.}
\end{figure}

As before, we quantify the short term prediction accuracy using the \gls{nrmse}. The results are shown in Figure~\ref{fig:L96_KS_shuffle_prediction_comparison}.
Immediately noticeable is the almost instant divergence of all \gls{sn} predictions. This is of course expected, considering the \gls{sn} neighborhood's assume a locally interacting system which the shuffled system is not. Furthermore, the \gls{nrmse} of the \gls{cc} and \gls{mi} neighborhoods is the combination of the \gls{nrmse} of the individual systems. This, again, makes sense due to the perfect separation between the \gls{l96} and the \gls{ks} system shown in Figure~\ref{fig:L96_KS_neighboorhoods}.
As for the \gls{ks} system, this results in the \gls{mi} \gls{sm} delivering significantly better results than the \gls{cc} \gls{sm}.

\begin{figure}[tph]
\centering
\includegraphics[width=\revtexFigWidth]{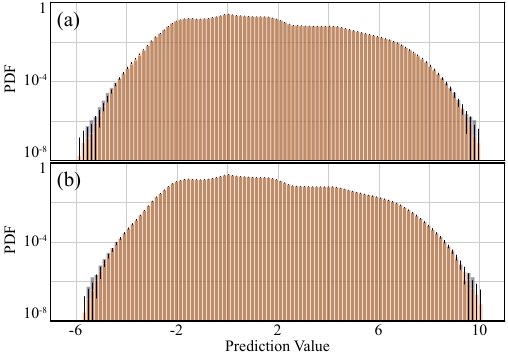}
\caption{\label{fig:L96_KS_pdfs}
\glspl{pdf} for the longterm predictions of the combined and shuffled KS-L96 system. The \glspl{pdf} are estimated via two superimposed 100 bin histograms of the simulated (gray) and the predicted (orange) data of the 
\textbf{(a)}
\gls{cc} or 
\textbf{(b)}
\gls{mi} Neighborhood.
Each entry in the histogram is generated by the prediction of a single variable value, with the dataset coming from the prediction of three 1000 Lyapunov time long sequences. The error bars represent the $1\sigma$ standard deviation of the predicted histogram bins.}
\end{figure}

For the long term statistical analysis we cannot use the \gls{mle} as it is not well defined due to the different time step sizes of the sub-systems. Therefore we look at the \glspl{pdf} to roughly estimate the system climate as also used in~\cite{Chattopadhyay2020}.
The details of the histogram generation are described in appendix~\ref{pdf_estimation}. \gls{cc} and \gls{mi} \glspl{pdf} are depicted in Figure~\ref{fig:L96_KS_pdfs}.

The predicted \glspl{pdf} show excellent agreement with the simulated data. Similarly to the \gls{mle} results, the worse short term predictive performance of the \gls{cc} \gls{sm} compared to the \gls{mi} \gls{sm} is not reflected in its climate reproduction quality.

\section{Conclusions} 
We have proposed and discussed a generalization of the concept of 
local states
in the sense of using 
similarity measures
derived from correlations among (instead of spatial distances between) time series. This offers a much more versatile approach for the prediction of high-dimensional complex systems.

First, 
generalized local states
can still make excellent predictions in the case of mixed systems, where 
local states
are doomed to fail. 
Here it is worth noticing that the perfect neighborhood separation we found in the combined \gls{ks}-\gls{l96} system (see Figure~\ref{fig:L96_KS_neighboorhoods}) suggests that 
generalized local states
can be used to separate mixed data sets and to thus infer different origins of a set of heterogeneous time series, for which the generating processes are unknown.

Second, prediction of high-dimensional systems remains feasible, when for a number or for all time series no spatial information is available. This is more and more the typical case in real world applications, when analyzing such heterogeneous data sets ranging from remote sensing data, financial data, social media data (e.g. Instagram, Twitter) to nowadays also infection rates during the COVID-19 pandemic, etc. and a combination of the aforementioned.       
Current research explores applications of 
generalized local states
based predictions in those cases.

\section{Acknowledgments}
We wish to acknowledge valuable discussions with D. Mohr, P. Huber, J. Aumeier, Y.Mabrouk, and J. Herteux.\\

\appendix
\section{CC Neighborhood Size}
\label{cc_neighborhood_size}

As discussed in the text, the \gls{cc} neighborhood sizes were necessarily chosen to be significantly larger than the \gls{sn} and \gls{mi} neighborhood sizes.

\begin{figure}[h]
\includegraphics[width=\revtexFigWidth]{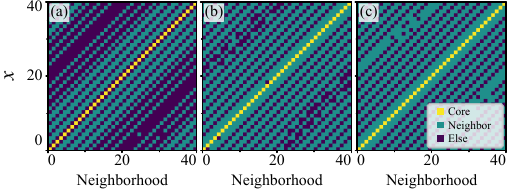}
\caption{\label{fig:L96_failed_CC_neighboorhoods} \gls{cc} neighborhoods of the \gls{l96} system. 
\textbf{(a)}
Total neighborhood size 19, the nearest neighbors of the core variables are not included in the neighborhood.
\textbf{(b)}
Neighborhood size 27. Only a couple nearest neighbors of the cores are missing. Nevertheless, predictions using these neighborhoods diverge regularly.
\textbf{(c)}
Neighborhood size 29. All nearest neighbors of the cores are included in the neighborhood. This is the smallest neighborhood size for which predictions consistently succeed. 
}
\end{figure}

This necessity is likely the result of the \gls{cc} \gls{sm} not recognizing the importance of the core's nearest neighbors in the \gls{l96} and \gls{ks} systems. As such, the \gls{cc} neighborhood size was increased to a total size of 29, the minimum where no predictions diverged and where the core's nearest neighbors were consistently included in its neighborhood. \gls{cc} neighborhoods of size 19, 27 and 29  for the \gls{l96} system are depicted in Figure~\ref{fig:L96_failed_CC_neighboorhoods}.

While the true importance of each variable regarding the prediction of another is of course unknown, 
at least for the locally interacting \gls{l96} and \gls{ks} systems studied here, the spatial nearest neighbors of the cores are critical for achieving a sensible prediction.

\section{Orbit Separation}
\label{orbit_separation}
Typically the most important Lyapunov exponent is considered to be the \gls{mle}, defined as the largest Lyapunov exponent of any given chaotic system. Its importance stems from the fact that it is intimately tied to the predictability of the system as, given a \gls{mle} $\Lambda_{\text{max}}$ two infinitesimally close trajectories in phase space, initially separated by the vector $\delta\boldsymbol{x}(t=0) $ diverge as
\begin{equation}
\label{eq:lyap_divergence}
\left|\delta\boldsymbol{x}(t)\right| \approx e^{\Lambda_{\text{max}} t} \left|\delta\boldsymbol{x}(t=0)\right|\,,
\end{equation}
where $t$ is the time since separation~\cite{Rosenstein1993}. 

To calculate the \gls{mle} we use Sprott's method of \gls{os}~\cite{sprott2003chaos}. 

By taking the logarithm and the average $\langle\cdot\rangle$ over many trajectory divergences in different parts of the chaotic attractor we find
\begin{equation}
\label{eq:lyap_divergence_fit}
\Lambda_{\text{max}} = \frac{1}{t_2 - t_1} \, \left\langle\frac{\log{\left|\delta\boldsymbol{x}(t_2)\right|}}{\log{\left|\delta\boldsymbol{x}(t_1)\right|}}\right\rangle\,.
\end{equation}
Note that for this equation to hold, we are only using the divergence data after transient effects have subsided but before the divergence size saturates due to it reaching the size of the chaotic attractor. 
From this, we can use a simple linear least squares fit to calculate the \gls{mle}. 

As described in the main text, we base the \gls{os} calculation on three 1000 Lyapunov time long term prediction datasets.
For each of the 100 realizations we choose 50 trajectory positions uniformly distributed in the first of the three long term datasets as starting point for the orbit separation 
at which we add normally distributed noise with standard deviation $\sigma_{\text{noise}} = 10^{-10} \sigma_{\text{r}}$ to the internal reservoir states. 
From this perturbed internal state, we let the reservoir predict 1500 time steps and compute the separation magnitude to the unperturbed predicted time series using the least squared fit of equation~\ref{eq:lyap_divergence_fit} as described above.

The \gls{mle} of the \gls{l96} and \gls{ks} simulations are computed analogously, using one $10^{5}$ time steps long dataset from which $10^{4}$ uniformly distributed starting positions for the trajectory divergence are chosen. The resulting \gls{mle}s are given in table~\ref{tab:system_lyap_properties}.
\begin{table}[h]
\begin{center}
\begin{tabular}{ |c|c|c|c|c| } 
 \hline
 System & $\Lambda_{\text{max}}$ & $\Delta t$ & $T_{L}$, & time steps per $T_{L}$, \\ 
\hline
 \gls{ks} & $0.049$ & $0.5$ & $20$ & $41$  \\ 
 \gls{l96} & $0.45$ & $0.05$  & $2.2$ & $45$ \\ 
 \hline
\end{tabular}
\caption{\glspl{mle} $\Lambda_{\text{max}}$, time step size $\Delta t$, Lyapunov time $T_{L}$, and the number of time steps per $T_{L}$ for the \gls{ks} and \gls{l96} systems.}
\label{tab:system_lyap_properties}
\end{center}
\end{table}

\section{\gls{pdf} Estimation}
\label{pdf_estimation}

\gls{pdf} estimation was done with the same data sets used to calculate the \gls{mle}.
The three resulting predictions and reference data sets from the simulation are treated as one single data set of 3000 Lyapunov times length respectively.
The resulting 
datasets are flattened and
then used to create the histogram.
The histograms themselves consist of 100 equally sized bins. To make comparison easy, they are chosen such that bin position and size for the simulated and predicted time series histograms are the same.

\bibliography{general_local_states_paper.bib, custom_references.bib}

\end{document}